\documentstyle[preprint,aps,version2]{revtex}
\begin{document}
\pagestyle{plain}

\begin{title}
Accessing directly the strange quark  content of the proton at HERA
\end{title}

\author{Wei Lu}

\begin{instit}
CCAST (World Laboratory), P.O. Box 8730, Beijing 100080, China

and Institute of High Energy Physics, Academia Sinica,
P.O. Box 918(4), Beijing 100039, China\footnote{Mailing address}
\end{instit}

\begin{abstract}
 We   investigate  a double-spin asymmetry for the
semi-inclusive $\Lambda$ hyperon
production in the  longitudinally  deep inelastic  lepton-proton
scattering,  the sign of which can  provide us with important information
about the strange quark helicity distribution in the proton.
On the basis of the interpretation of the  longitudinal
 deep inelastic lepton-nucleon scattering data as
a  negative strange quark polarization in the proton and  the preliminary
results on the measurement of  the  longitudinal $\Lambda$ polarization
at the $Z$ resonance in electron-positron annihilation,
we predict a minus sign for the suggested observable.  The
experimental condition required for  our suggestion is
met by the  HERA facilities, so the asymmetry  considered
can be  measured by the  HERMES experiments  at HERA in the near future.
\end{abstract}

\newpage

Since the announcement of the famous European Muon Collaboration
(EMC) experiment results \cite{EMC}, the extensive interest in the physics
community  has been attracted  \cite{review} by the proton spin structure.
Besides the EMC data, the ensuing  experiments \cite{ensue}
also indicate that the strange quarks and antiquarks  in the
proton possess a net negative polarization opposite to
the proton spin. However, other  possible interpretations
such as the polarized glue in the proton \cite{gluon}
have been competing with  this  allegation.
Therefore, it is imperative  to invent some machineries
to access independently the strange  contents of the proton.

Considering the self-analyzing  property of the $\Lambda$
hyperon, a group of authors  have discussed
the possibility to  use the $\Lambda$ polarization as
a strange quark polarimeter to  detect the strange quark
polarization in the proton.  Based on an extrapolation
of the present knowledge of polarized  structure functions [1,3],
Ellis et al. \cite{ex}  suggested a model for the
strangenees of the proton,  in which  a valence quark core
with the  naive quark model spin content  is accompanied  by
a spin-triplet $\bar s s$ pair.  Furthermore,  the strange
antiquark  is supposed to be negatively polarized, motivated
by  chiral dynamics, and likely the strange quark, motivated
by $^3P_0$ quark condensation in the vacuum.
In such a model, Alberg, Ellis, and Kharzeev \cite{AEK}
pointed out that the measurement of target spin depolarization
parameter in the $\bar p p \to \bar \Lambda \Lambda$
may construct a test of the
polarization state of the strange quarks in the proton.
Following this line, Ellis, Kharzeev and Kontzinian  \cite{EKK}
made predictions for the polarization of the
$\Lambda$ hyperons semi-inclusively
detected in the  {\it target fragmentation region}.
At the same time, Lu and Ma  \cite{lm} investigated
the polarization of  the $\Lambda$ particles
produced in the {\it current fragmentation region}  within  the
framework of the quark-parton model. Assuming an
unpolarized lepton beam and a longitudinally polarized proton target,
they found that the sign of the $\Lambda$ polarization
in the current fragmentation region can
supply us with important information about the
strange quark polarization in the proton.

 At present, the  most promising deeply inelastic scattering
experiments are  being done at  DESY HERA, where
both  the lepton  beam and nucleon target are longitudinally
polarized. Therefore, it is desirable to  search for some
observables  accessible at the HERA facilities
to probe the strange quark spin in the proton.
The  HERA experiments are performed
\cite{tech} by keeping the beam polarization unchanged and reversing  the
target polarization as required. In this Letter, we completely
conform to such experimental conditions and investigate the
implication of the possible negative strange quark polarization
to the semi-inclusively detected $\Lambda$ particles.
Our findings are  positive, i.e., there exists
a  spin asymmetry  for the semi-inclusive $\Lambda$
hyperon production when both the lepton beam and proton target
are longitudinally polarized.  The sign  of this quantity
is contingent on that of the strange quark polarization in the proton.

More generally, we consider the semi-inclusive $\Lambda$ production
by the longitudinally  deep inelastic lepton-nucleon  scattering
$$
l(p_l, s_l) +N (p_N,s_N) \to l(p_l^\prime)
+\Lambda(p_\Lambda, s_\Lambda) +X,
$$
where the  particle momenta and  covariant spin vectors are self-explanatory.
As a first approximation, we adopt the one-photon
exchange approximation, in which  the proton structure
is probed  by a virtual photon  of momentum $q=p_l-p^\prime_l$.
Then, the cross section   is  related to the Lorentz contraction between
the leptonic tensor and hadronic tensor.

 As usual, the leptonic  tensor is taken as
\begin{eqnarray}
L^{\mu\nu}(q,p_l,s_l)&=&\frac{q^2}{2} g^{\mu\nu}
+2p_l^\mu p_l^\nu -p_l^\mu q^\nu -q^\mu p_l^\nu
+2im_l \varepsilon^{\mu\nu\tau\rho}
q_\tau s_{l\rho},
\end{eqnarray}
while the hadronic tensor  is defined
\begin{eqnarray}
W^{\mu\nu}(q,p_N,s_N,p_\Lambda,s_\Lambda)&\equiv &
\frac{1}{4\pi}
\sum \limits_X
\int d ^4 \xi \exp(i q \cdot \xi )
\nonumber \\
& & \times
\langle   p_N,s_N|j^\mu (0)|\Lambda(p_\Lambda, s_\Lambda),X\rangle
\langle  \Lambda(p_\Lambda, s_\Lambda),X|j^\nu(\xi)|p_N,s_N\rangle,
\end{eqnarray}
where the electromagnetic current is defined as
$j_\mu=\sum\limits_a e_a \bar\psi_a\gamma_\mu\psi_a$
with  $a$ the quark flavor index and $e_a$
the quark charge in unit of the electron charge.
We normalize the spin vector  as
$s\cdot s=-1$ for the pure state of
a spin-half fermion. In the  forthcoming presentation,
the  longitudinal spin four-vector $s$  is related to the  particle
 helicity  $h$ via
\begin{equation}
{\rm limit}_{m\to 0}~ ms^\mu =hp^\mu,
\end{equation}
where $m$ is the fermion mass and $p^\mu$ the momentum.

We adopt the conventional scalar variables:
\begin{equation}
x_B=\frac{-q^2}{2 p_N\cdot q} ,~ y=\frac{p_N\cdot q}{p_N\cdot p_l},~
z=\frac{p_N\cdot p_\Lambda}{p_N\cdot q}.
\end{equation}
Correspondingly, the cross section can  be written as
\begin{equation}
\frac{d  \sigma (s_l,s_N,s_\Lambda) }{d  x_B d  y  d  z
d ^2 {\bf  p}_{\Lambda\perp}}
=
\frac{ \alpha^2 y  }{8 \pi^2 z Q^4} L_{\mu\nu}(q,p_l,s_l)
W^{\mu\nu} (q, p_N, s_N, p_\Lambda, s_\Lambda),
\label{5}
\end{equation}
where  $Q=\sqrt{-q^2}$ and ${\bf p}_{\Lambda\perp}$ is the components of the
transverse $\Lambda$ momentum relative to the axis of the
quark  fragmentation jet.

The hadronic tensor contains all  the information about
the proton structure and $\Lambda$ hyperon production.
Because of the lack of methods  to treat nonperturbative
effects,  the  general strategy so far is to  factorize \cite{CSS}
the hadronic tensor into the  long- and short-distance
parts.  The long-distance
matrix elements encode  the information on
the proton structure and the  hyperon
production by parton  hadronization  whereas the
short-distance coefficients describes the hard partonic interaction.
We will  work  at the leading twist factorization, which is
equivalent to the quark-parton model prescription without
including  any higher-order effects, so the corresponding nonperturbative
matrix elements can be interpreted as the quark distribution
and fragmentation functions in the quark parton model.
For our purpose to elucidate the main physics,
it is enough to adopt such a lowest-order approximation.

 At the leading order and leading twist,  only the lowest-order
diagram shown in Fig. 1  makes contributions to the  hadronic tensor:
\begin{eqnarray}
\int W^{\mu\nu} (q, p_N, s_N, p_\Lambda, s_\Lambda)
\frac{d^3 p_\Lambda}{2 E_\Lambda (2\pi)^3}
&=&\frac{1}{4\pi}\int \frac{d^4 p_\Lambda}{(2\pi)^4}
(2\pi)\delta(p^2_\Lambda-M^2_\Lambda)
\nonumber\\
& &\times
\int \frac{d^4 k}{(2\pi)^4}
\sum\limits_a e^2_a {\rm Tr}\left[
T^a_N(k,p_N,s_N)\gamma_\mu T^a_\Lambda (k+q, p_\Lambda,s_\Lambda)\gamma_\nu
\right],
\end{eqnarray}
where  two nonperturbative matrices (in the Dirac space)
 are defined as
\begin{equation}
T^a_N(k,p_N,s_N)_{\alpha\beta}=\int d^4 \xi \exp(ik\cdot \xi)
\langle  p_N,s_N|\bar\psi^a_\beta(0)\psi^a_\alpha(\xi)|p_N,s_N\rangle ,
\end{equation}
\begin{equation}
T^a_\Lambda(k,p_\Lambda,s_\Lambda)_{\alpha\beta}=\sum\limits_X
\int d^4 \xi \exp(-ik\cdot \xi)
\langle  0|\psi^a_\alpha(0)|\Lambda (p_\Lambda,s_\Lambda),X\rangle
\langle  \Lambda (p_\Lambda,s_\Lambda),X|\bar\psi^a_\beta(\xi)|0\rangle .
\end{equation}
For the  lowest-order diagram,  its leading  twist contributions
can be extracted most efficiently  by  use of the collinear
expansion technique \cite{EFP}, i.e., carrying
 out an expansion of parton momenta with respect to their
components collinear with the corresponding hadron momenta.
In this work, we restrict ourselves with
the $\Lambda$ production in the current fragmentation
region so that  the effects  of the $\Lambda$ hyperon mass
can be ignored and  the transverse $\Lambda$ momentum can be
safely integrated out.   Undergoing  the standard procedure\cite{JJ1,JJ2}
to separate the nonperturbative matrix elements from the hard
partonic interaction part, we obtain
the following   leading twist factorization results:
\begin{eqnarray}
\int W^{\mu\nu} (q, p_N, h_N, p_\Lambda, h_\Lambda)
 d^2 {\bf p}_{\Lambda\perp}
&=&\frac{1}{2zp_N\cdot q}
\sum\limits_a e^2_a [f^a_1(x_B)\hat f_1^a (z)
(-p_N\cdot p_\Lambda g^{\mu\nu} +
p^\mu_N p^\nu_\Lambda +p^\mu_\Lambda p^\nu_N )
\nonumber\\
& & +ih_N {\sl g}_1^a (x_B) \hat f_1^a (z)
 \varepsilon^{\mu\nu\lambda\sigma}q_\lambda p_{N\sigma}
+ih_\Lambda f_1^a (x_B) \hat {\sl g}_1^a (z)
 \varepsilon^{\mu\nu\lambda\sigma}q_\lambda p_{\Lambda\sigma}
\nonumber \\
& & + h_N h_\Lambda {\sl g}^a_1(x_B) \hat {\sl g}^a_1(z)
(-p_N\cdot p_\Lambda g^{\mu\nu} +
p^\mu_N p^\nu_\Lambda +p^\mu_\Lambda p^\nu_N )].
\label{9}
\end{eqnarray}
where $f_1(x)$ and  ${\sl g}_1(x)$ are the quark momentum
distribution and quark helicity distribution in the
proton, $\hat f_1(x)$ and  $\hat {\sl g}_1(x)$ are
the spin-independent and longitudinal spin-dependent
quark fragmentation functions for inclusive $\Lambda$ production.
(We follow Jaffe and Ji's notations  about the quark distribution
functions \cite{JJ1}  and  fragmentation functions \cite{JJ2}.)
As a matter of fact,  eq. (\ref{9})
can also be derived from  the  quark-parton  model.

Substituting eqs. (1) and (\ref{9}) into (\ref{5}),  we  deduce the
following expression for the cross section:
\begin{eqnarray}
\frac{d  \sigma (h_l,h_N,h_\Lambda) }{d  x_B d  y  d  z }
&=&
\frac{ \alpha^2   }{16 \pi^2 y z Q^2}
\sum\limits_a [ (y^2-2y+2) f^a_1(x_B)\hat f_1^a (z))
\nonumber \\
& & +h_Nh_\Lambda(y^2-2y+2) {\sl g}^a_1(x_B) \hat {\sl g}^a_1(z)
+h_l h_N y (2-y) {\sl g}^a_1(x_B)
+h_l h_\Lambda y (2-y)\hat {\sl g}^a_1(z)]. \label{10}
\end{eqnarray}

For the HERA experiments, both the  lepton beam and nucleon
target  are in their  helicity states.  We consider the
polarization of the detected $\Lambda$ hyperons, which
is defined as
\begin{equation}
P_\Lambda(h_l,h_N)\equiv
\frac
{d\sigma(h_l,h_N,+)-d\sigma(h_l,h_N,-)}
{d\sigma(h_l,h_N,+)+d\sigma(h_l,h_N,-)},
\end{equation}
where $\pm$ are shorthand for $\pm \frac{1}{2}$.
{}From eq. (\ref{10}), one can obtain
\begin{equation}
P_\Lambda(h_l,h_N)=
\frac
{\sum\limits_a e^2_a [h_N(y^2-2y+2)
 {\sl g}^a_1(x_B) \hat {\sl g}^a_1(z)+h_l y(2-y) \hat {\sl g}^a_1(z)]}
{\sum\limits_a e^2_a [
(y^2-2y+2)f^a_1(x_B)\hat f^a_1(z)+h_l h_N y(2-y) {\sl g}^a_1(x_B)]}.
\end{equation}
Obviously, the $\Lambda$ polarization  has
two sources:  the  spin transfer from the lepton beam and
that   from  the nucleon target, respectively.

Keeping in mind the valence quark configuration of the $\Lambda$ hyperon,
we may assume $a$ $priori$ that the $\Lambda$ particle is predominantly
produced by the current strange quark fragmentation. Then,
the flavor summation  can be dropped in the above formula and
correspondingly we consider only the contributions pertinent to
the strange quark. As Burkardt and Jaffe \cite{BJ} have  discussed,
the measurement of the longitudinal $\Lambda$ polarization around the
$Z$ resonance in electron-positron annihilation
can allow the determination
of the $s\to \Lambda$ fragmentation functions, both $\hat f^s_1(z)$
and $\hat {\sl g}^s_1 (z)$. Since the LEP I  collider was operated
at  the $Z$ resonance and HERA is presently colliding
28 GeV  electrons on 820 GeV protons,  the corresponding
processes are  commonly believed to fall into the  reign
in which perturbative QCD works. Once
the $s\to \Lambda$  fragmentation functions are obtained  by
analyzing LEP-I data, one can  evolve them to the HERA energy scale
by the Altarelli-Parisi equations \cite{AP}.
With such inputs, the measurement of the $\Lambda$
polarization  at HERA will make possible an
independent extraction of  the strange quark
helicity distribution ${\sl g}^s_1(x)$.  However,  this
will not be an easy work because of the experimental  complexities.

 For  our understanding of the strangeness in the
proton, the  sign of  the net strange quark polarization may
be more important than  its precise $x$-dependence.
Our finding is that
the double-spin asymmetry
\begin{equation}  \label{defi}
A(h_l)\equiv\frac
{d\sigma(h_l,+,+)-d\sigma(h_l,+,-)-d\sigma(h_l,-,+)+d\sigma(h_l,-,-)}
{d\sigma(h_l,+,+)+d\sigma(h_l,+,-)+d\sigma(h_l,-,+)+d\sigma(h_l,-,-)}
\end{equation}
is sensitive to  the sign of ${\sl g}_1^s(x)$.  To see this
point,  one can insert eq. (\ref{10}) into (\ref{defi}), obtaining
 the following simple formula
\begin{equation}
A=
\frac
{ {\sl g}^s_1(x_B) \hat {\sl g}^s_1(z)}
{f^s_1(x_B)\hat f^s_1(z)},
\end{equation}
in which we again take into  account the contributions relevant
to the strange quark only.
Therefore,  although this result is subject to radiative corrections
 and higher-twist effects,  one can at least anticipate that
the measurement of $A(h_l)$  allow the determination of the sign
of ${\sl g}^s_1(x)$.

   The measurement of $A(h_l)$ needs to monitor the
polarization  of the $\Lambda$ hyperon,  so $d\sigma
(h_l, h_N,h_\Lambda)$ is  not  directly measurable.
However,  the asymmetry $A(h_l)$  can be accessed
experimentally.  To show this, we cast $A(h_l)$ into the form
\begin{equation}
A(h_l) = \sum_{h_N} {\rm sign} (h_N) D(h_l, h_N) P(h_l, h_N),
\end{equation}
where
\begin{equation}
D(h_l,h_N)\equiv \frac
{\sum\limits_{h_\Lambda} d\sigma(h_l,h_N,h_\Lambda)}
{\sum\limits_{h^\prime_N, h_\Lambda} d\sigma(h_l,h^\prime_N, h_\Lambda)}.
 \end{equation}
So, $A(h_l)$ can be schematically thought of as being  a
$D(h_l,h_N)$-weighted ``asymmetry" of the $\Lambda$
polarization when the helicity of the target nucleon is reversed.
Obviously, $D(h_l,h_N)$ can be  measured  by  controlling
the initial-state beam and target polarizations in the
deep inelastic scattering  and  semi-inclusively detecting
a $\Lambda$ hyperon  in the  current fragmentation region.
Therefore,  the asymmetry $A(h_l)$ is an  observable
experimentally accessible.  Hopefully,  its measurement can be
implemented by the future  HERMES experiments \cite{tech}
at  HERA
as well as by the future  HELP experiment \cite{CERN} at CERN.

The preliminary  results about the
longitudinal $\Lambda$ polarization at  the $Z$ resonance
at LEP-I  have already
been existent \cite{pre}, which imply that the
longitudinal spin-dependent $s\to \Lambda$ fragmentation function
is positive. Although  the data are  subjected to refinement,
one can be confident that there will not be qualitative changes.
If the interpretation is accepted  of the  longitudinal
deeply inelastic scattering data  as a  net negative
strange quark polarization  in the proton,  we  can
make a prediction that the  considered  double-spin asymmetry
is negative. The experimental condition required
for this spin asymmetry  happens to be met at HERA,
where  one fixes the lepton beam polarization  but
reverse the target polarization as required so as to
measure  the  longitudinal spin-dependent   quark distributions.
Considering  the large statistics needed,  one can at least
determine the sign of the strange quark polarization in the proton.

In conclusion, we worked out a leading twist  factorized
expression for the hadronic tensor for the
semi-inclusive $\Lambda$ production by the
longitudinally  deep  inelastic scattering of leptons off nucleons.
Furthermore, we proposed  an
observable $A(h_l)$,
which  is simply related to the strange quark
distribution functions in the nucleon and the $s\to \Lambda$
fragmentation functions.  Provided that the involved
fragmentation functions are  precisely  measured
at LEP in the near future,  the measurement of the
suggested quantity  at HERA will allow  the determination of
the strange quark  helicity distribution in the nucleon.
Considering the large statistics needed in such experiments,
we  point out  that even  the  accurate measurements cannot be
done on $A(h_l)$,  its measured sign
can  still  supply us  with  very useful information about the
strange  quark polarization in the proton.

 The author thanks Professor P. S\"oding for useful
conversations as well as  for supplying a copy of Ref. \cite{tech}.

\centerline{\bf \large Figure Caption}
Figure 1. The lowest-order diagram  contributing to the hadronic tensor
for the semi-inclusive $\Lambda$ hyperon production  in the
deeply inelastic scattering of leptons off nucleons.

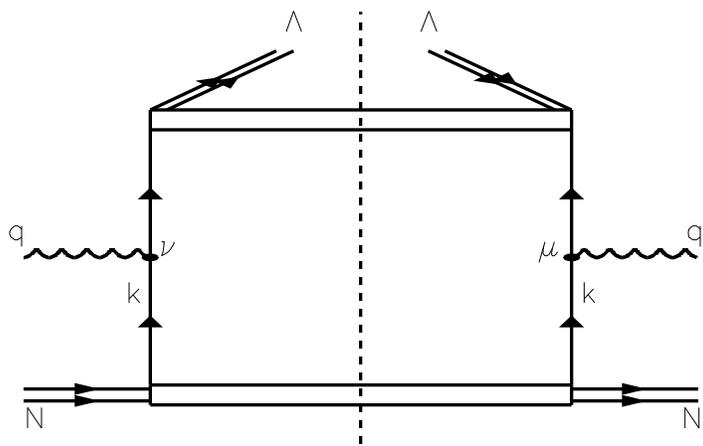
\end{document}